\documentclass[preprint,12pt]{elsarticle}

\usepackage{amssymb}
\usepackage{xspace}
\usepackage{hepparticles}
\usepackage{hepnames}

\usepackage{hyperref}
\usepackage{listings}
\usepackage{xcolor}
\usepackage{array}
\usepackage{longtable}
\newcolumntype{L}[1]{>{\raggedright\let\newline\\\arraybackslash\hspace{0pt}}m{#1}}
\newcolumntype{C}[1]{>{\centering\let\newline\\\arraybackslash\hspace{0pt}}m{#1}}
\newcolumntype{R}[1]{>{\raggedleft\let\newline\\\arraybackslash\hspace{0pt}}m{#1}}

\def \heplike{{\texttt{HEPLike}}\xspace}
\def \chisq{\ensuremath{\chi^2}\xspace}

\def \yaml{\texttt{YAML}\xspace}

\def \cls{\ensuremath{\rm CL_s}\xspace}

\def \clspvalue{$\rm CL_s$/p-value\xspace}

\def \pvalue{p-value\xspace}
\def \hepdata{HEPData\xspace}

\def \cpp{\texttt{C++}\xspace}

\def \pvalue{p-value\xspace}
\def \root{\texttt{ROOT}\xspace}
\def \bib{\texttt{BiBtex}\xspace}
\def \ttree{\texttt{TTree}\xspace}

\def \hldata{{\texttt{HL\_Data}}\xspace}

\def \hllimit{{\texttt{HL\_Limit}}\xspace}
\def \hlgauss{{\texttt{HL\_Gaussian}}\xspace}
\def \hlbifgauss{{\texttt{HL\_BifurGaussian}}\xspace}
\def \hlndimgauss{{\texttt{HL\_nDimGaussian}}\xspace}
\def \hlndimbifurgauss{{\texttt{HL\_nDimBifurGaussian}}\xspace}

\def \hlproflike{{\texttt{HL\_ProfLikelihood}}\xspace}
\def \hlndimlike{{\texttt{HL\_nDimLikelihood}}\xspace}
\def \hlexpdata{{\texttt{HL\_ExpPoints}}\xspace}

\def \github{\texttt{github}\xspace}
\def \lhcb{LHCb\xspace}
\def \cms{CMS\xspace}

\def \atlas{ATLAS\xspace}

\newcommand{\secref}[1]{Sec.~\ref{#1}} 
\newcommand{\figref}[1]{Fig.~\ref{#1}} 
\newcommand{\Eqref}[1]{Eq.~\ref{#1}} 
\newcommand{\tabref}[1]{Tab.~\ref{#1}}

\newcounter{bla}

\journal{Computer Physics Communications}

\begin{document}

\begin{frontmatter}



\title{\texttt{HEPLike}: an open source framework for experimental likelihood evaluation}


\author[a]{Jihyun Bhom\corref{author}}
\author[a]{Marcin Chrzaszcz\corref{author}}

\cortext[author] {Corresponding author.\\\textit{E-mail address:} jihyun.bhom@ifj.edu.pl, mchrzasz@cern.ch}
\address[a]{Henryk Niewodniczanski Institute of Nuclear Physics Polish Academy of Sciences, Krak´ow, Poland}

\begin{abstract}

We present a computer framework to store and evaluate likelihoods coming from High Energy Physics experiments. Due to its flexibility it can be interfaced with existing fitting codes and allows to uniform the interpretation of the experimental results among users. The code is provided with large open database, which contains the experimental measurements. The code is of use for users who perform phenomenological studies, global fits or experimental averages. 
\end{abstract}


\end{frontmatter}



{\bf PROGRAM SUMMARY/NEW VERSION PROGRAM SUMMARY}

\begin{small}
\noindent
{\em Program Title: HEPLike}                                          \\
{\em Licensing provisions(please choose one): GPLv3 }                                   \\
{\em Programming language: C++}                                   \\

{\em Supplementary material:}                                 \\
{\em Journal reference of previous version: FIRST VERSION OF PROGRAM}                  \\

{\em Nature of problem(approx. 50-250 words):  Provide a uniform way of store, share and evaluate experimental likelihoods in a proper statistical manner. The code can be easily interfaced with existing global fitting codes. In addition a large database with the measurements is published. The program targets users who perform in their scientific work: phenomenological studies, global fits or measurements averages. The HEPLike has been created for FlavBit project[1], which was used to perform several analysis[2,3] and here we present an updated version, which can be used in standalone mode.
}\\
{\em Solution method(approx. 50-250 words): C++ code that evaluates the statistical properties of the measurements without user intervention. The large open database is provided as well. The measurements are stored in \yaml files allowing for easy readability and extensions. }\\

\end{small}

\clearpage

\section{Introduction}
\label{sec:introduction}

In the High Energy Physics (HEP) the experimental measurements are performed by several collaborations, which measure various different observables. The experimental results are presented in various ways; some being as simple as a measurement with an Gaussian error, some more complicated as multiple correlated measurements with asymmetric errors or in some places even a full likelihood function is being published. To make things more complicated in some cases multiple representations of the same measurement are being published. All of this makes it hard to directly use and compare various different results. It also leaves a room for misinterpreting the results by theorists, which use these inputs to their studies. It happens that the asymmetric errors are being symmetrized, instead of using the full likelihood only central value with approximated asymmetric error is being used. 

The High Energy Physics Likelihoods (\heplike) is a computer program that allows to store and share the likelihoods of various measured quantities. The published code can be useful for users performing phenomenological studies using experimental results, global fitting collaborations or experimental averages. Thanks to its structure it is easy to be interface with existing codes. It simplifies the work of people as instead of looking up the appropriate measurement and coding up their own likelihood they can download the database of measurements and choose the one they need. Furthermore, it shifts the burden of constructing the proper likelihood functions back  to the experimentalists, which performed the measurement at the first place and are clearly the most appropriate people to handle this task.

The computer code described in this paper is written in \cpp, making it useful for majority of fitting programs available on the market~\cite{Athron:2017ard,Costa:2017gup,Bechtle:2004pc,Mahmoudi:2007gd,Feldmann:2018kqr,Kumar:2018kmr}. The library can be used in both the \chisq and likelihood fits. Moreover, it contains a statistical module with useful functions that can be used in the statistical analysis. Besides the computer code a database with the likelihoods is being published. The measurements are stored in the \yaml files making them easy to read by both the machine and human. This database can be easily extended by adding new \yaml files if new measurement becomes available. With the software we provide useful utilities, which allows to perform searches inside the database, create \bib containing publications, which have been in the fit, etc.

The paper is organized as follows: in \secref{sec:stat} construction of the likelihood functions is presented. \secref{sec:implementation} explains the detailed code implementations and data storage, while \secref{sec:installation} describes how to install and use \heplike software.

\section{Likelihood constructions}
\label{sec:stat}

In this section we will present how likelihoods in \heplike are stored and constructed. Each measurement is stored in separate \yaml file. There are several ways collaborations published their results depending on the measurements itself:
\begin{itemize}
\item Upper limits,
\item Single measurement with symmetric uncertainty,
\item Single measurement with asymmetric uncertainty,
\item Multiple measurements with symmetric uncertainty,
\item Multiple measurements with asymmetric uncertainty,
\item One dimensional likelihood function,
\item n-dimensional likelihood function.
\end{itemize}

In addition, there is growing interest from the community that the experimental collaborations instead of only the results of the analysis publish also the dataset that has been used to obtain the result. For this future occasion we have also implement a way that this data can be directly used in the fits.

Each of these cases has a different module of \heplike that is designed to evaluate the likelihood functions. In this section we will present the statistical treatment of the above cases and the modules that are responsible for their evaluation. Each of the \yaml files is required to have the following information (here for example we use the  $R_{\PKstar}$ measurement~\cite{Aaij:2017vbb}):

\newcounter{codecounter}
\begin{lstlisting}[escapeinside={(*}{*)}, numbers=right]
BibCite: Aaij:2017vbb                                                             
BibEntry: '@article{Aaij:2017vbb,                                                 
      author         = "Aaij, R. and others",                                     
      title          = "{Test of lepton universality
                         with $B^{0} \rightarrow      
                        K^{*0}\ell^{+}\ell^{-}$ decays}",                         
      collaboration  = "LHCb",                                                    
      journal        = "JHEP",                                                    
      volume         = "08",                                                      
      year           = "2017",                                                    
      pages          = "055",                                                     
      doi            = "10.1007/JHEP08(2017)055",                                 
      eprint         = "1705.05802",                                              
      archivePrefix  = "arXiv",                                                   
      primaryClass   = "hep-ex",                                                  
      reportNumber   = "LHCB-PAPER-2017-013, 
                        CERN-EP-2017-100",                   
      SLACcitation   = "%%CITATION = ARXIV:1705.05802;%%"                         
      }                                                                           
      '                                                              
DOI: 10.1007/JHEP08(2017)055                                                      
Process: R_{Kstar^{*}}                                                              
FileName: RKstar.yaml                                                                                                                                  
Name: RKstar                                                                      
Source: HEPDATA                                                                   
SubmissionYear: 2017
PublicationYear: 2018
Arxiv: 1705.05802
Collaborations: LHCb
Kinematics: q2>1.1 && q2<6. 
HLAuthor: Gal Anonim
HLEmail: gal.anonim@ifj.edu.pl
HLType: HL_ProfLikelihood
\end{lstlisting}

\noindent The above informations are used to store the information relevant for the bookkeeping. For instance the entries \texttt{BibCite} and \texttt{BibEntry} correspond to the information that are used to generate a \bib citation file with the measurements that have been used in the studies. The \texttt{DOI} corresponds to the digital object identifier of the publication. The \texttt{Decay} defines the process that has been studied. It can also be replaced by the \texttt{Process} entry. The \texttt{Name} is a unique name of this measurement type. If the measurement gets updated with more data or by other collaboration the \texttt{Name} entry in the new \yaml file should be the same as in the old one. \texttt{Source} entry corresponds to the source of the measurement. This can be either a \hepdata or the collaboration itself. The \texttt{SubmissionYear} (\texttt{PublicationYear}) refers to the year of appearance (publication) of the result. The \texttt{Arxiv} codes the Arxiv number, while the  \texttt{Collaborations} stores the information which experimental collaboration has performed the measurement. Finally, the \texttt{Kinematics} stores additional information about the kinematic region that has been measured. The \texttt{HLAuthor} and \texttt{HLEmail} encode the information about the \yaml file author and his email in case user needs further information about the encoded measurement. Last but not least the entry \texttt{HLType} contains the information about which \heplike object should be used to read the file.

Reading of this content in the \yaml is implemented in the \texttt{HL\_Data} class. All other classes that construct the likelihood functions inherit from this class its capabilities. Please note that if the information is missing in the \yaml file the program will omit reading this entry. The only exception is the \texttt{FileName}, which is mandatory. If a user wants to be notified by the program that some informations are missing the \texttt{HL\_debug\_yaml} variable has to be set to \texttt{true} (default value is \texttt{false}).

\subsection{Upper limits}
\label{sec:UL}

\begin{figure}[htb]
\center
\includegraphics[width=0.45\textwidth]{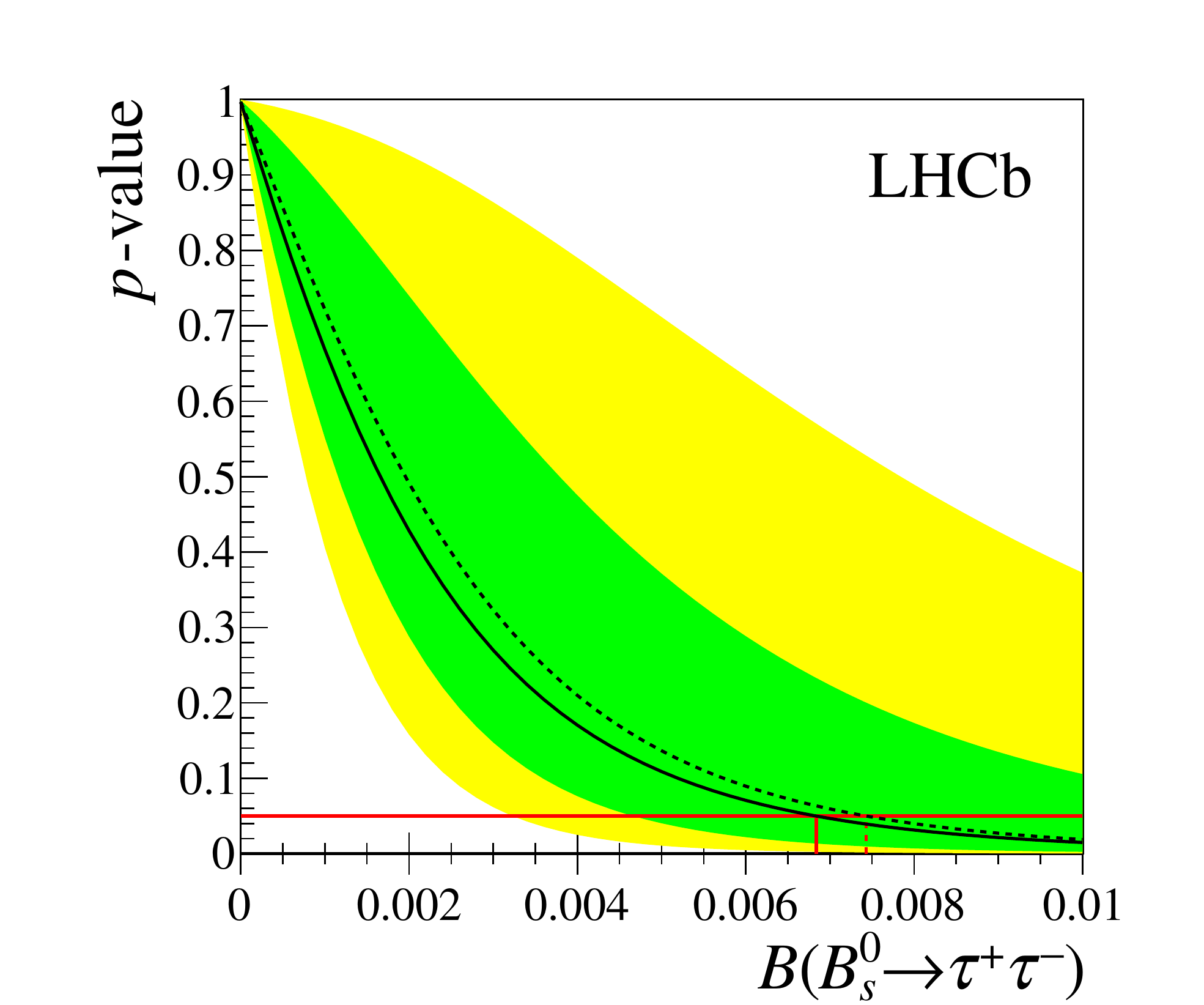}
\includegraphics[width=0.45\textwidth]{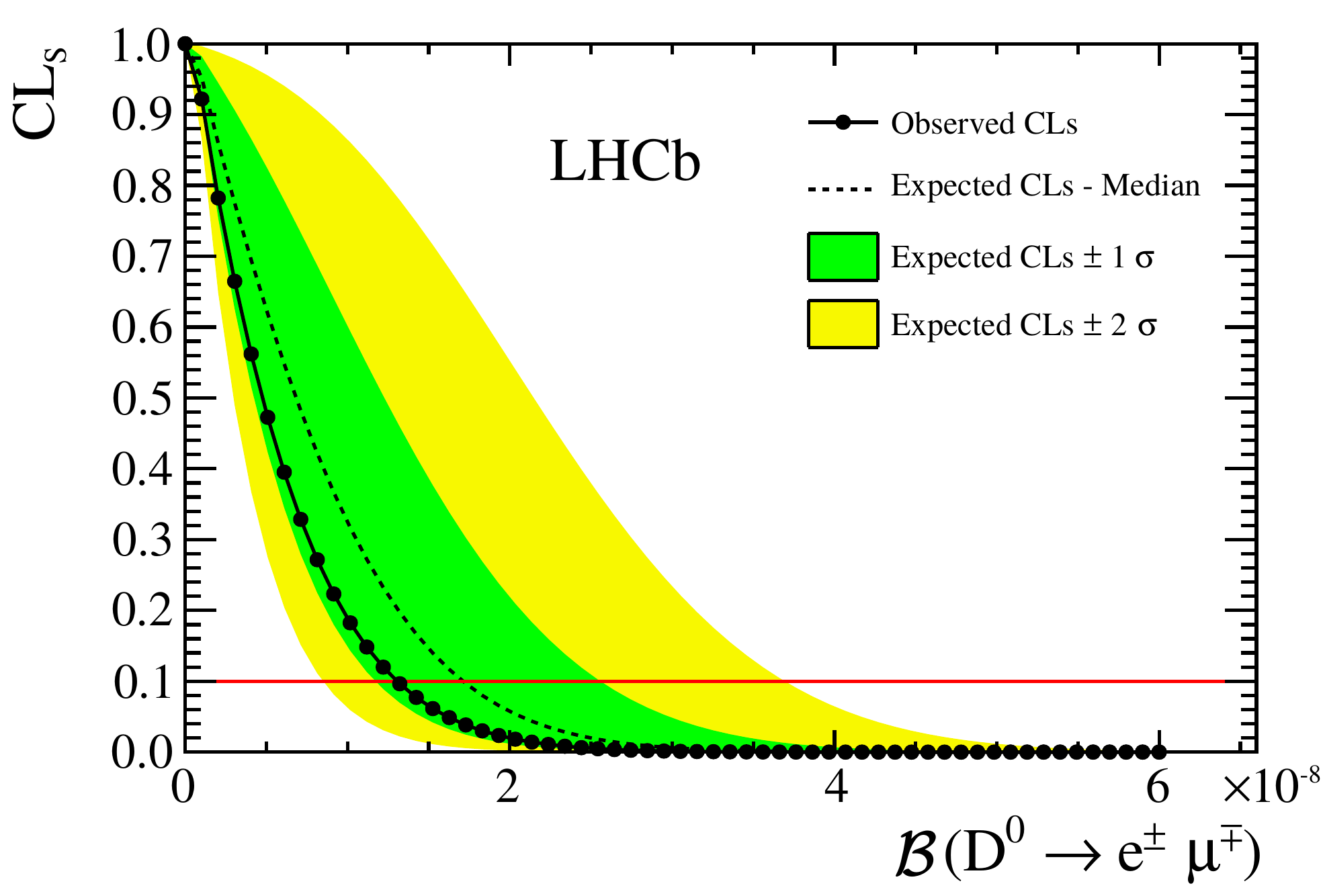}

\caption{Example of \pvalue scans for the $\PBs \to \Ptauon \APtauon$~\cite{Aaij:2017xqt} (left) and $\PD \to \Pe \Pmu$~\cite{Aaij:2015qmj} (right). Please note that the \cls~value can be interpreted as p-value as explained in~\cite{Read:451614}. The black line corresponds to the observed \clspvalue. \label{fig:UL}}
\end{figure}

In case where the measurement did not observe a significant access of signal candidates the collaborations usually report an upper limit on the measured quantity. Commonly $90\%$ or $95\%$ upper limits are quoted. Experiments use various statistical approaches to compute this limits. It can be the \cls method~\cite{Aaij:2015qmj}, Feldman--Cousins~\cite{Feldman:1997qc} or some variation of Bayesian methods~\cite{Rover:2011zq}. Publication of only an upper limits does not provide enough information to use the result in global fits. However, nowadays experiments besides the aforementioned upper limits publish a full p-value scans. Examples of such scans are shown in \figref{fig:UL}. The plots are usually available in digital format, which allows the information to be extracted and used in computer program.

In \heplike a class \hllimit is responsible for handling this type of measurements. It reads the \yaml file that contains the standard information about the measurement (see~\secref{sec:stat} for details). The additional information of the observed \clspvalue is stored in the \yaml file in the following way\footnote{Please note that the besides this information the previous information from \secref{sec:stat} should be included.}:
\begin{lstlisting}[escapeinside={(*}{*)}, numbers=right]
Cls:                          
- [0.0, 1.0]                  
- [1.0e-10, 0.977091694706]   
- [2.0e-10, 0.954375824297]   
- [3.0e-10, 0.93200355343]    
- [4.0e-10, 0.910630700546]   
- [5.0e-10, 0.889382721809]   
\end{lstlisting}

\noindent The \texttt{Cls} can be replaced in the \yaml file by \texttt{p-value} as they correspond to the same information. The first number in each array is the value of tested hypothesis (for example branching fraction), while the second is the corresponding \clspvalue. These values are then interpreted using a \chisq distribution with one degree of freedom:

\begin{equation}
pdf(x) = \frac{1}{2^{1/2} \Gamma(1/2)} x^{1/2 -1} e ^{-x/2},
\end{equation}
which had the cumulative distribution function defined as:
\begin{equation}
cdf(x)=\frac{1}{\Gamma(1/2) } \gamma(1/2,x/2).
\end{equation}
In the above equations the $\Gamma(x)$ and $\gamma(k,x)$ correspond to Gamma and incomplete gamma functions.
By revering the $cdf(x)$ one can obtain the \chisq value:
\begin{equation}
\chisq=cdf^{-1}(1-p),
\end{equation}
where p corresponds to the p-value of a given x hypothesis. This \chisq can be then translated to the log-likelihood via Wilks theorem~\cite{wilks1938}:
\begin{equation}
-\log(\mathcal{L})= \frac{1}{2}\chisq, \label{eq:wilks}
\end{equation}
where the $\mathcal{L}$ is the likelihood. The user can choose if he wants to obtain the \chisq, likelihood or a log-likelihood value of a given hypothesis.

\subsection{Single measurement with symmetric uncertainties}
\label{sec:gauss}

The simplest case of a published experimental result is a single value with a symmetric uncertainty. This is for example a typical result of an PDG of HFLAV average~\cite{PhysRevD.98.030001,Amhis:2016xyh}. The measurement is coded in the \yaml file as:
\begin{lstlisting}[escapeinside={(*}{*)}, numbers=right]
Observables:                         
- [ "Br_A2BCZ", 0.1, 0.05, 0.01 ]    
\end{lstlisting}

The first argument in the array "Br\_A2BCZ" corresponds to the observable name. Then the first number corresponds to the measured central value. The 2nd and the 3rd number are the statistical and systematic uncertainties. In case where only one uncertainty is available the 3rd number should be omitted and it will be automatically set to 0 in the software. We have decided to keep the plural \texttt{Observables}  to be more uniform in case where more observables are measured.

The module responsible for reading this \yaml file is called \hlgauss, it calculates the \chisq for an $x$ hypothesis:
\begin{equation}
\chisq = \frac{(x_{obs}-x)^{2}}{  \sigma_{stat}^{2}+ \sigma_{syst}^{2} },
\end{equation}
where the $x_{obs}$ correspond to the measured central value in the \yaml file and the $\sigma_{stat}$ and $\sigma_{syst}$ are the statistical and systematic uncertainties respectively. This can be the again translated to the likelihood and log-likelihood value using \Eqref{eq:wilks}.

\subsection{Single measurement with asymmetric uncertainties}
\label{sec:bifurgauss}

A simple extension of the Gaussian uncertainty is when an asymmetric uncertainty is reported. This type of measurements although less frequent appear in the literature. The publication in this case reports the central value and two uncertainties: $\sigma_+$ and $\sigma_-$, which correspond to the right (for values larger than the measured central value) and left (for values smaller than the measured central value) uncertainty. In \heplike we have created a \hlbifgauss class, which reads the following entry in the \yaml file:
\begin{lstlisting}[escapeinside={(*}{*)}, numbers=right]
Observables:                                        
- [ "Br_A2BCZ", 0.1, 0.05, -0.06,  0.01, -0.02 ]    
\end{lstlisting}

The first argument is again the name of the observable and the second one is its central value. The third and fourth arguments correspond to the statistical $\sigma_+$ and $\sigma_-$ uncertainties, while the fifth and sixth to the systematical $\sigma_+$ and $\sigma_-$ uncertainties. It is important to keep the minus sign before the left side uncertainties. The code will indicate the error in case of missing sign. In some cases the systematical uncertainty is reported to be symmetric. In such case the last number can be omitted in the \yaml entry.

In the literature there exist number of ways to interpret asymmetric uncertainties~\cite{Barlow:2003sg}. We have chosen the most commonly used one which is the so-called bifurcated Gaussian:
\begin{align}
\chisq = 
\begin{cases}
    \frac{(x_{obs}-x)^{2}}{  \sigma_{+}^{2}} ,& \text{if } x \geq x_{obs}\\
    \frac{(x_{obs}-x)^{2}}{  \sigma_{-}^{2}} ,& \text{if } x < x_{obs},\\
\end{cases}
\end{align}
where the  $\sigma_{\pm}^{2}$ is the sum of squared statistical and systematic uncertainties for right/left case. Once \chisq is calculated it can be translated to the log-likelihood using \Eqref{eq:wilks}.

\subsection{Multiple measurements with symmetric uncertainties}
\label{sec:ndimgauss}

Nowadays the most common are simultaneous measurements of various quantities, which are correlated between each other. For instance cross section measurements in different kinematic bins, or measurements of angular coefficients in heavy mesons decays. In \heplike the class responsible for handling these cases is called \hlndimgauss. It reads the following information from the \yaml file:
\begin{lstlisting}[escapeinside={(*}{*)}, numbers=right]
Observables:                                 
- [ "BR1", 0.1, 0.02]                        
- [ "BR2", 0.2, 0.01, 0.01]                  
- [ "BR3", 0.4, 0.04]                        
Correlation:                                 
-  [ "BR1", "BR2", "BR3"]                    
-  [ 1. ,  0.2 ,  0   ]                      
-  [ 0.2,   1.,   0. ]                       
-  [ 0 ,   0.,  1.  ]                        
\end{lstlisting}                                  
The information in the ``Observables'' entry is exactly the same as in the \hlgauss class. Please note that similarly to the previous class the systematic uncertainty is not mandatory and in case if it is not provided the code will treat it as 0. The next entry in the \yaml file is the ``Correlation'', which encodes the correlation matrix. The first ''row'' is the names of the variables it is important to keep the same order of variables as in the ``Observables'' entry. The \hlndimgauss evaluates the \chisq in the following way:
\begin{align}
\chisq =  V^{T} {\rm Cov}^{-1} V,\label{eq:chi2ndim}
\end{align}
where V is a column matrix, which is the difference between the measured and the tested i-th observable value. The $\rm Cov$ is a square matrix, constructed from the correlation matrix (${\rm Corr}$): ${\rm Cov}_{i,j}={\rm Corr}_{i,j} \sigma_i \sigma_j$.

Often a user does not want to use the full set of measured quantities but just their subset. In this case a function \texttt{Restrict(vector<string>)} can be used. By passing in a form of vector the list of observables to be used, the program will create new smaller covariance matrix, which will be used to evaluate the \chisq. In a similar manner the \chisq can be translated to the likelihood and log-likelihood value by~\Eqref{eq:wilks}.

\subsection{Multiple measurements with asymmetric uncertainties}
\label{sec:ndimbifgauss}

More complicated case is when multiple correlated measurements are reported with asymmetric uncertainties. The case is similar to the one discussed in~\secref{sec:bifurgauss} and same statistic comments apply in this case. The \yaml file encoding such a measurement will contain the following entries:
\begin{lstlisting}[escapeinside={(*}{*)}, numbers=right]
Observables:                                                     
- [ "BR1", 0.1, +0.02, -0.01, 0.02]                              
- [ "BR2", 0.2, +0.01, -0.05, +0.03, -0.02]                      
- [ "BR3", 0.3, +0.04, -0.03, 0.05]                              
Correlation:                                                     
-  [ "BR1", "BR2", "BR3"]                                        
-  [ 1. ,  0.1 ,  0.2   ]                                        
-  [ 0.1,   1.,   0.1 ]                                          
-  [ 0.2 ,   0.1,  1.  ]                                                 
\end{lstlisting}

The meaning of the ``Observables'' entry is the same as in the previous class (cf.~\secref{sec:bifurgauss}) and the ``Correlation'' encodes the same information as in the \hlndimgauss class (cf..~\secref{sec:ndimgauss}). The rules about the minus sign and symmetric systematic uncertainty are the same as in case of the \hlbifgauss (cf.~\secref{sec:bifurgauss}). The difference arises when one evaluates the \chisq, namely the $\rm cov$ matrix is constructed depending if $\sigma_+$ and $\sigma_-$ uncertainty is relevant:
\begin{align}
{\rm Cov}_{i,j}=
\begin{cases}
{\rm Corr}_{i,j}~\sigma^{i}_+ \sigma^{j}_+, & \text{if } x^i \geq x^i_{obs} \text{ and }  x^j \geq x^j_{obs} \\
{\rm Corr}_{i,j}~\sigma^{i}_+ \sigma^{j}_-, & \text{if } x^i \geq x^i_{obs} \text{ and }  x^j < x^j_{obs} \\
{\rm Corr}_{i,j}~\sigma^{i}_- \sigma^{j}_+, & \text{if } x^i < x^i_{obs} \text{ and }  x^j \geq x^j_{obs} \\
{\rm Corr}_{i,j}~\sigma^{i}_- \sigma^{j}_-, & \text{if } x^i < x^i_{obs} \text{ and }  x^j < x^j_{obs} \\
\end{cases}
\end{align}

The obtained $\rm Cov$ matrix is then used to calculate the \chisq using~\Eqref{eq:chi2ndim}. The rest follows the same procedure as described in~\secref{sec:ndimgauss}.

\subsection{One dimensional likelihood function}
\label{sec:onedimlikelihood}

\begin{figure}[htb]
\center
\includegraphics[width=0.45\textwidth]{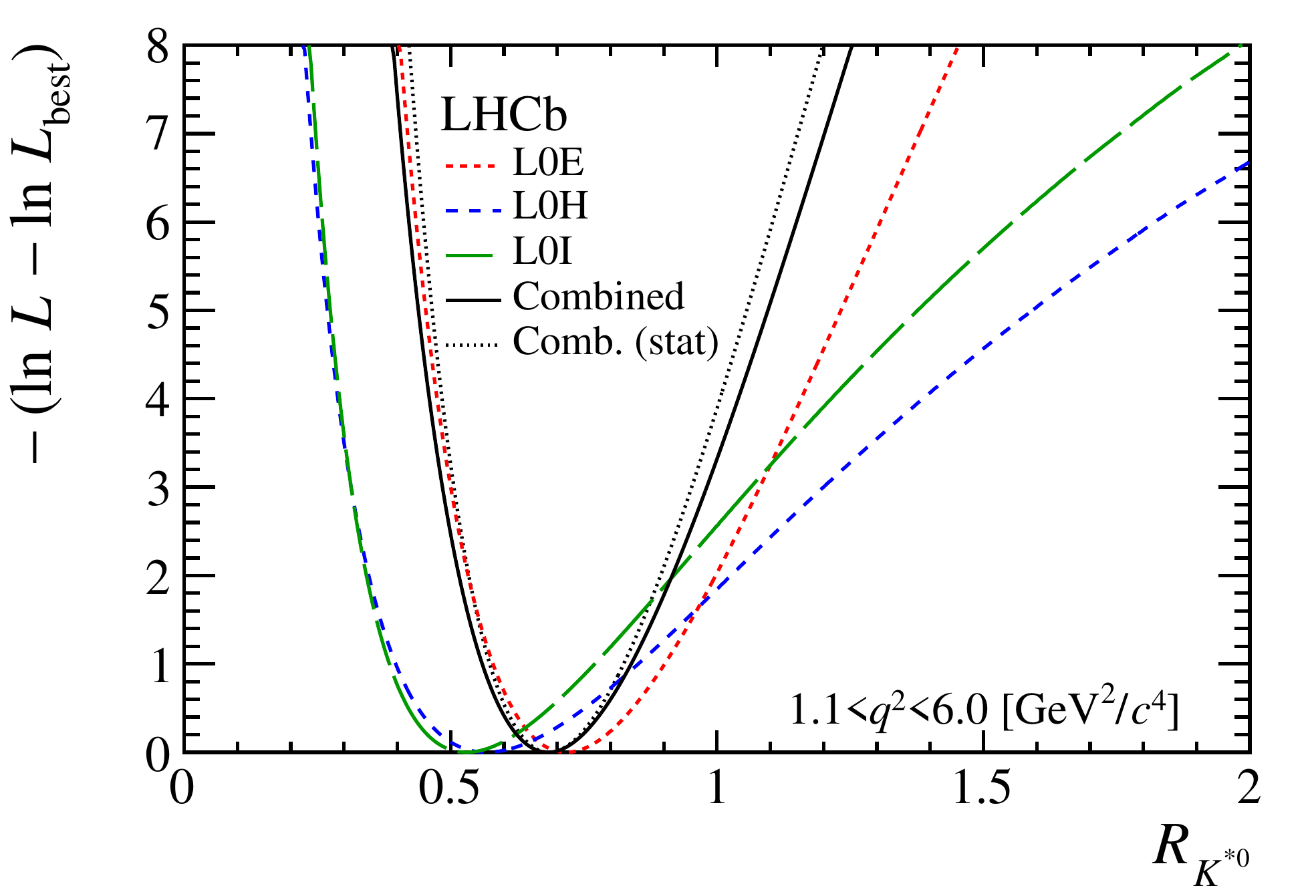}
\includegraphics[width=0.45\textwidth]{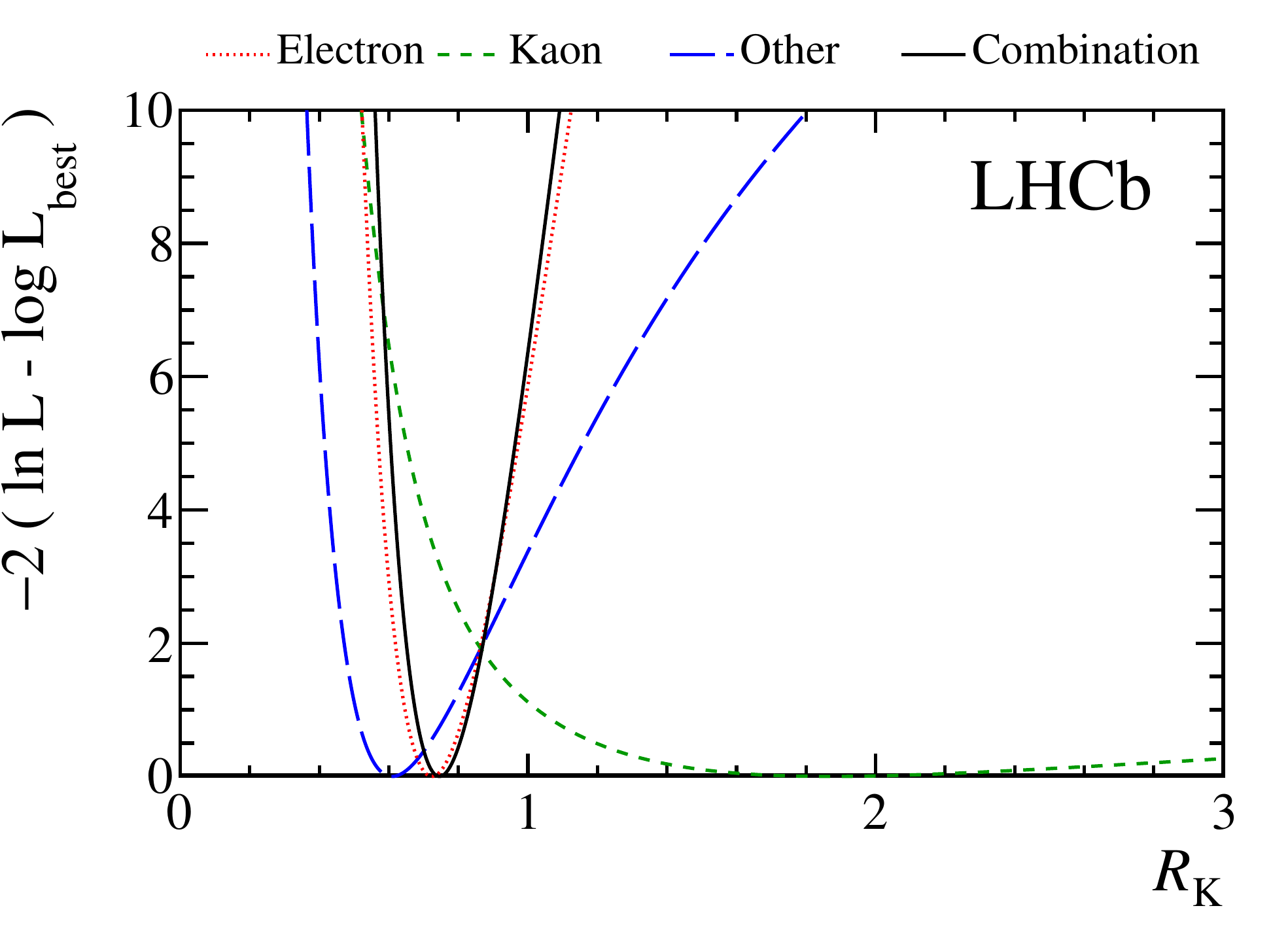}

\caption{Examples of published one-dimensional likelihoods in the Lepton Universality Violation of the  $\PB \to \PKstar \ell \ell$~\cite{Aaij:2017vbb} (left) and  $\PB \to \PK \ell \ell$~\cite{Aaij:2014ora} (right).
 \label{fig:OneDimLikelihood}}
\end{figure}

The best way a result can be published is by providing the (log-)likelihood function. This type of results are more and more common in the literature. The most easy is the one-dimensional likelihood scans as can be presented in form of a figure, which examples are shown in~\figref{fig:OneDimLikelihood}.

The biggest advantage of publishing the results in this form is its completeness. The (log-)likelihood curve contains all the information about all the non-Gaussian effects and incorporates the systematic uncertainties. The technical problem is how to publish such information. Usually plots are published in the \texttt{pdf} or \texttt{png} formats which makes them hard to be used. Since experiments are mostly using \root~\cite{Antcheva:2009zz} framework the plots are saved also in the \texttt{C} format, which contains the points in the form of arrays. This of course makes the points accessible however it is not easy to automate retrieving this data from the \texttt{C} file. The best solution is provided by the \texttt{HEPData} portal~\cite{Maguire:2017ypu}. It allows to download the data in a user preferred format. In \heplike we have chosen to use the \root format by default, in which the data points are saved in the form of a \texttt{TGraph} object, which is also the way experimentalists like to store this information. In the \yaml file we specify the path of the \root in the following way:
\begin{lstlisting}[escapeinside={(*}{*)}, numbers=right]
ROOTData: data/HEPData-ins1599846-v1-Table_1.root
TGraphPath: "Table 1/Graph1D_y1"                                               
\end{lstlisting} 

\noindent The \texttt{ROOTData} encodes the location of the \root file, while the \texttt{TGraphPath} encodes the location of the \texttt{TGraph} object in that \root file. In \heplike the class \hlproflike is responsible for reading and encoding this likelihood. The value of the log-likelihood can be ten translated again into the \chisq with \Eqref{eq:wilks}.

\subsection{n-dimensional likelihood function}
\label{sec:ndimlikelihood}

\begin{figure}[htb]
\center
\includegraphics[width=0.45\textwidth]{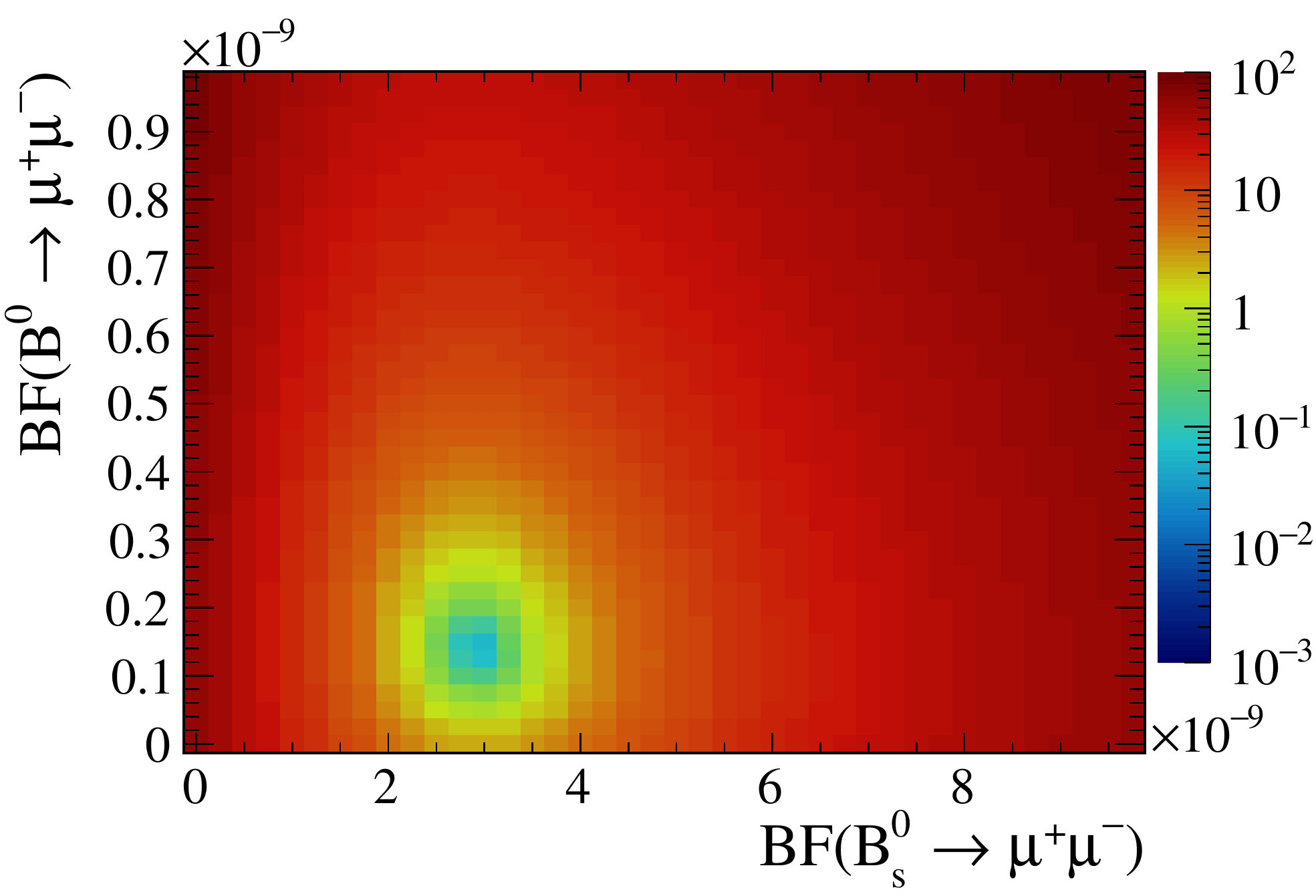}
\includegraphics[width=0.45\textwidth]{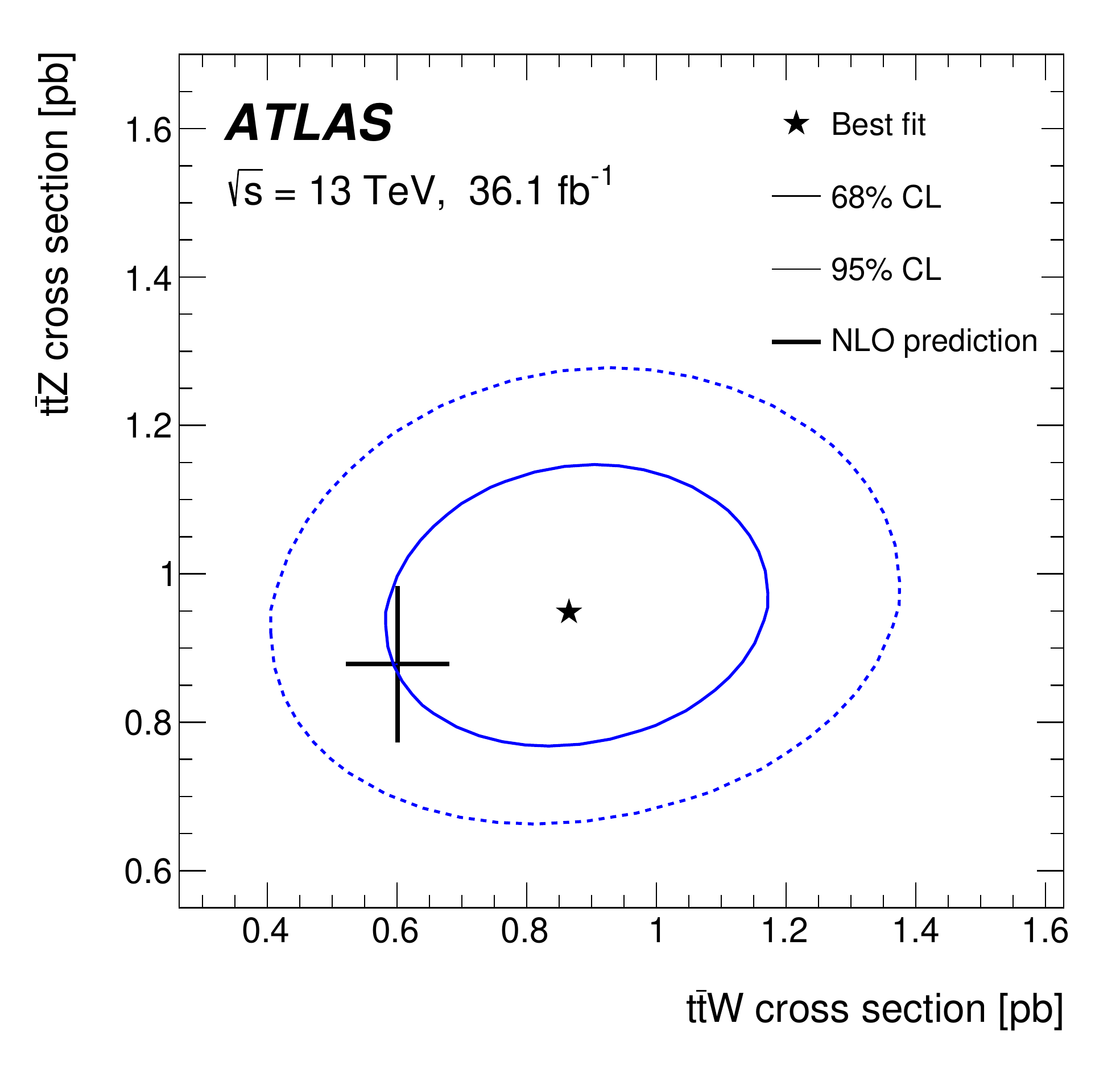}

\caption{Examples of published two-dimensional likelihoods. The $\mathcal{B}(\PBs \to \mu \mu)$ vs $\mathcal{B}(\PBd \to \mu \mu)$ likelihood~\cite{Aaij:2017vad} (left) and  $\sigma(\Ptop\APtop \PZ)$ vs $\sigma(\Ptop\APtop \PW)$ likelihood~\cite{Aaboud:2019njj} (right).
 \label{fig:nDimLikelihood}}
\end{figure}

The natural extension of one dimensional likelihood is an n-dim  likelihood, where $n \geq 2$. Currently experimental collaborations publish only 2-dim likelihood functions (cf.~\figref{fig:nDimLikelihood}).

The natural way of encoding such information is a histogram: \texttt{TH2D} or \texttt{TH3D} and we have chosen this way to store this information. The corresponding entry in the \yaml file looks as following:
\begin{lstlisting}[escapeinside={(*}{*)}, numbers=right]
ROOTData: data/LHCb/RD/Bs2mumu_5fb/histB2mumu.root
TH2Path: "h_2DScan"                                            
\end{lstlisting} 
Similar to the one dimensional likelihood (\secref{sec:onedimlikelihood}) the \texttt{ROOTData} encodes the location of the \root file, while the \texttt{TH2Path}(\texttt{TH3Path}) encodes the location of the \texttt{TH2D}(\texttt{TH3D}) object. In the long run the community will have to address the question how to publish higher dimensional likelihoods and this module (\hlndimlike) will have to be extended for such use cases.

\subsection{Fits to experimental data}
\label{sec:Expdata}

It is possible that in the future experimental collaborations besides the results will made the datasets public. The procedure and the form in which the data should be published is not decided and there is an ongoing debate if the published data should correspond to the raw detector data, the final selected points used in the analysis or something between? Clearly publishing a raw data is problematic, as people outside the collaboration do not have the necessary knowledge about the calibration and efficiency correction procedures or data taking conditions. The most useful way to publish the dataset is to allow the experimentalists to perform all the selection, all the necessary efficiency corrections and publish the final dataset that has been used for analysis. This would allow the theory community to use the dataset directly in their fits without knowing the technicalities about the experimental data analysis. For this case in \heplike we have implemented such a class \hlexpdata.

The data are stored in the \ttree structure located in the \root file. The \yaml file encodes this information in form:
\begin{lstlisting}[escapeinside={(*}{*)}, numbers=right]
ROOTData: data/toy/data.root                                        
TTreePath: t                                                        
Observables:                                                        
- [ x ]            
- [ y ]
- [ z ]                                              
Weight: w                                                                                                      
\end{lstlisting} 
where the \texttt{ROOTData} points to the \root file and the \texttt{TTreePath} stores the information of the \ttree location inside the \root file. It is assumed that the experiments will provide all the corrections in form of event-by-event weights. The name of the weight inside the \ttree is encoded in the \texttt{Weight} entry. In general the data points are elements of $\mathcal{R}^n$ vector space, which coordinates are stored in the \texttt{Observables} entry. 

The only thing that user needs to provide to the \hlexpdata object is a pointer to the function to be fitted. The function should have a form: \texttt{double (*fun)(vector<double> par , vector<double> point)}, where the 
\texttt{par} vector encodes the parameters that want to be fitted and the point corresponds to a data point. The \hlexpdata will then evaluate the likelihood:
\begin{equation}
\mathcal{L}(\omega)= f(\textbf{x} | \omega)^{w(\textbf{x} )} \label{eq:wlikelihood}
\end{equation}
for the whole dataset. In the above the $\textbf{x}$ correspondents to the n-dimensional point, $\omega$ denotes the parameters that want to be fitted \texttt{par}, and $f$ denotes the fitting function (\texttt{fun}). The \heplike does not provide a minimalizer or a scanner tool as it is not purpose of this type of software. It has to be interfaced with proper scanner tool for example~\cite{Athron:2017ard}. Again the user can decide if he/she prefers to perform a \chisq or log-likelihood fit.

The biggest advantage of such format is the compatibility with the experimental analysis. Experimentalist can in principle publish as well the function that they have used to fit this data and therefore a theorists reproduce the experimental result and start where the experimentalists finished.

\section{Code implementation}
\label{sec:implementation}

In this section we will discuss the implementation of the code used to create likelihoods discussed in \secref{sec:stat}. The code is build in several classes:
\begin{itemize}
\item \hldata: base class from which other classes inherit their base functionality.
\item \hllimit: class that handles the upper limit measurements.
\item \hlgauss: class that handles measurements with Gaussian uncertainty.
\item \hlbifgauss: class that handles measurements with asymmetric uncertainty.
\item \hlndimgauss: class that handles measurements with n-dimensional Gaussian uncertainties.
\item \hlndimbifurgauss: class that handles measurements with n-dimensional asymmetric uncertainties.
\item \hlproflike: class that handles measurements with one-dimensional likelihood function.
\item \hlndimlike: class that handles measurements with 2(3)-dimensional likelihood function.
\item \hlexpdata: class that allows to perform the fits to experimental datasets.
\end{itemize}

\begin{figure}[htb]
\includegraphics[width=0.99\textwidth]{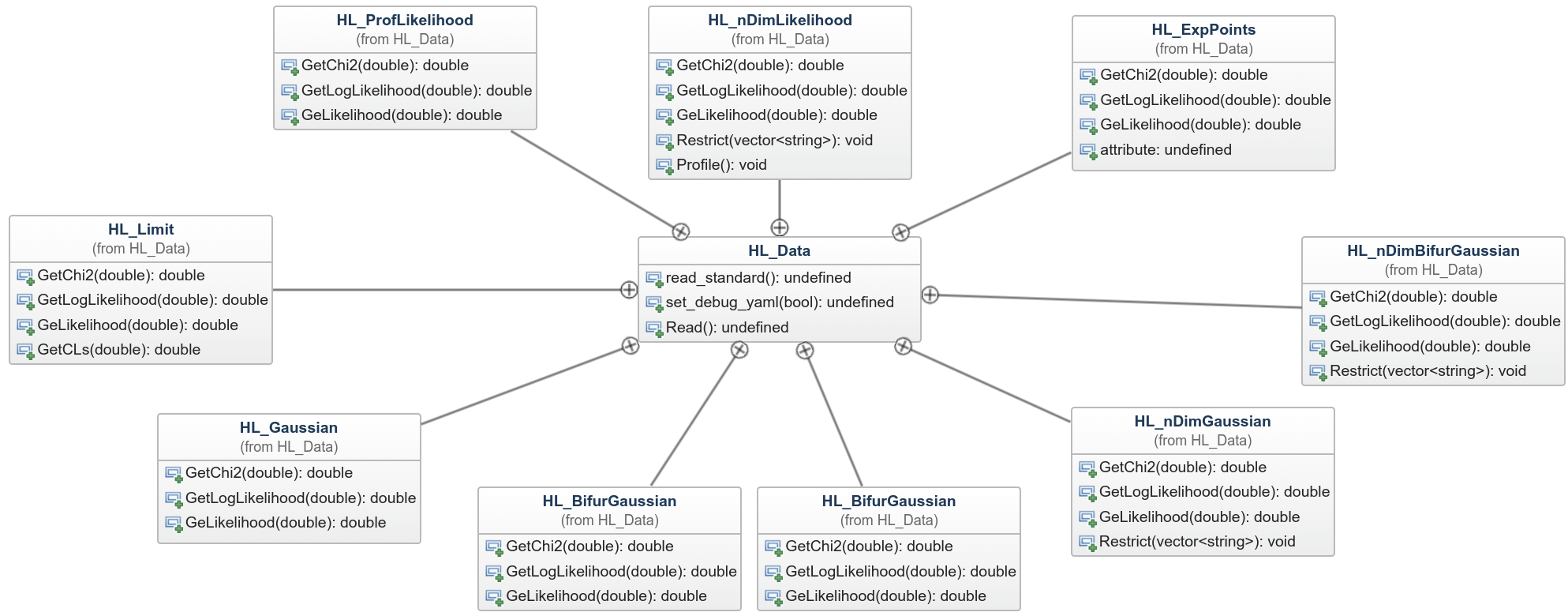}
\caption{Diagram of class inheritance of the \heplike package.\label{fig:diagram}}
\end{figure}

In \tabref{tab:functions} we present the functionality of these classes. In addition we present the hierarchy of the structure of the class inheritance in \figref{fig:diagram}. 

\begin{longtable}{|c|L{8cm}|}
\caption{Functions available in the \heplike software. \label{tab:functions}}\\
\hline
\textbf{Function} & \textbf{Description} \\
\hline
\endfirsthead
\multicolumn{2}{c}%
{\tablename\ \thetable\ -- \textit{Continued from previous page}} \\
\hline
\textbf{Function} & \textbf{Description} \\
\hline
\endhead
\endfoot
\hline
  \texttt{HL\_Data()} &  Constructor of the \hldata class.\\  \hline
   \texttt{HL\_Data(string)} & Constructor of the \hldata class. The argument that is taken by constructor is the path for the \yaml file encoding the measurement.\\  \hline
  \texttt{HL\_Limit()} &  Constructor of the \hllimit class.\\  \hline
   \texttt{HL\_Limit(string)} & Constructor of the \hllimit class. The argument that is taken by constructor is the path for the \yaml file encoding the measurement.\\  \hline
    \texttt{HL\_Gaussian()} &   Constructor of the \hlgauss class.\\  \hline
   \texttt{HL\_Gaussian(string)} & Constructor of the \hlgauss class. The argument that is taken by constructor is the path for the \yaml file encoding the measurement.\\  \hline 
  \texttt{HL\_BifurGaussian()} &   Constructor of the \hlbifgauss class.\\  \hline
   \texttt{HL\_BifurGaussian(string)} &  Constructor of the \hlgauss class. The argument that is taken by constructor is the path for the \yaml file encoding the measurement.\\  \hline 
    \texttt{HL\_nDimGaussian()} &  Constructor of the \hlndimgauss class.\\  \hline
   \texttt{HL\_nDimGaussian(string )} & Constructor of the \hlndimgauss class. The argument that is taken by constructor is the path for the \yaml file encoding the measurement.\\  \hline  
    \texttt{HL\_nDimBifurGaussian()} &  Constructor of the \hlndimbifurgauss class.\\  \hline
   \texttt{HL\_nDimBifurGaussian(string)} & Constructor of the \hlndimbifurgauss class. The argument that is taken by constructor is the path for the \yaml file encoding the measurement.\\  \hline    
     \texttt{HL\_ProfLikelihood()} & Constructor of the \hlproflike class.\\  \hline
   \texttt{HL\_ProfLikelihood(string)} & Constructor of the \hlproflike class. The argument that is taken by constructor is the path for the \yaml file encoding the measurement.\\  \hline  
      \texttt{HL\_nDimLikelihood()} &  Constructor of the \hlndimlike class.\\  \hline
   \texttt{HL\_ProfLikelihood(string)} & Constructor of the \hlndimlike class. The argument that is taken by constructor is the path for the \yaml file encoding the measurement.\\  \hline  
      \texttt{HL\_ExpPoints()} &  Constructor of the  \hlexpdata class.\\  \hline
   \texttt{HL\_ExpPoints(string)} & Constructor of the  \hlexpdata class. The argument that is taken by constructor is the path for the \yaml file encoding the measurement.\\  \hline   
 \texttt{read\_standard()} & Function that reads the general information about the measurement from the \yaml file.  \\ \hline
 \texttt{set\_debug\_yaml(bool)}   &  Function that enables debugging the \yaml file. By default the debugging is switched off and can be switched on by passing a \texttt{true} bool argument to this function. Debugging will print a message that for a given information in the \yaml file is missing.  \\ \hline
  \texttt{Read()} & Function reading the \yaml file. The  function \\ \hline
      \texttt{GetChi2(double)} & Function that returns the \chisq value for a given point (passed to the function as double). Function is available for all classes besides \hldata. \\ \hline  
    \texttt{GetLogLikelihood(double)} &  Function that returns the log-likelihood value for a given point (passed to the function as double). Function is available for all classes besides \hldata. \\ \hline  
  \texttt{GetLikelihood(double)} &  Function that returns the likelihood value for a given point (passed to the function as double). Function is available for all classes besides \hldata. \\ \hline  
    \texttt{GetCLs(double)} & Function that returns  \cls or \pvalue for a given point (passed to the function as double). The function is a member of the \hllimit class. \\  \hline  
   \texttt{Restrict(vector<string>)}  &  Function that restricts number of observables from the \yaml file. Function is a member of the \hlndimgauss, \hlndimbifurgauss and \hlndimlike classes.   \\  \hline  
     \texttt{InitData()}  & Function of \hlexpdata class that reads to the memory the data from the \ttree object.  \\  \hline  
          \texttt{Profile()}  & Function of \hlndimlike class that creates the profile log-likelihood projections. \\  \hline  
     \texttt{SetFun()}  & Function of \hlexpdata class, that sets the pointer to the function to be fitted.  \\  \hline  

\end{longtable}

\section{Installation and usage}
\label{sec:installation}

In this chapter we will present the requirements and installation for the \heplike package. The software is distributed via the \github site: \url{https://github.com/mchrzasz/HEPLike}.

In order to compile \heplike the following packages (and the minimal version) needs to be installed:
\begin{itemize}
\item \texttt{git}
\item \texttt{cmake}, 2.8
\item \texttt{yaml-cpp}, 1.58.0
\item \texttt{gsl}, 2.1
\item \texttt{Boost}, 1.58.0
\item \texttt{ROOT}, 6.08
\end{itemize}

The compilation is done in the following way:
\begin{lstlisting}[escapeinside={(*}{*)}, numbers=right]
cd <instalation dir>
git clone  https://github.com/mchrzasz/HEPLike.git       
cd HEPLike
mkdir build
cd build
cmake ..
make   
\end{lstlisting} 
In the above the \texttt{make} can be replaced with \texttt{make -jN}, where N is the number of threads that user wants to be used for compilation. Please note that in case of non standard installation of some packages one might have to provide \texttt{cmake} with a proper path to the library. After successful compilation a \texttt{libHEPLike.a} and  \texttt{libHEPLike.so}libraries will be created in the \texttt{build} directory.

The \heplike is provided with seven examples:
\begin{itemize}
\item \texttt{Br\_example.cc}: example program  showing the usage of the \hlgauss class.
\item \texttt{BrBifurGaussian\_example.cc}: example program showing the usage of the \hlbifgauss class.
\item \texttt{Data\_Fit\_example.cc}: example program showing the usage of the \hlexpdata class.
\item \texttt{Limit\_example.cc}: example program showing the usage of the \hllimit class.
\item \texttt{Ndim\_BifurGaussian\_example.cc}: example program showing the usage of the \hlndimbifurgauss class.
\item \texttt{Ndim\_Gaussian.cc}: example program showing the usage of the \hlndimgauss class.
\item \texttt{Ndim\_Likelihood\_example.cc}: example program showing the usage of the \hlndimlike class.
\item \texttt{ProfLikelihood\_example.cc}: example program showing the usage of the \hlproflike class.
\end{itemize}

To compile them a proper variable has to be set during the cmake stage:
\begin{lstlisting}[escapeinside={(*}{*)}, numbers=right]
  cd build
  cmake -DEXECUTABLE=TRUE ..
  make
\end{lstlisting}

After the compilation in the \texttt{build} directory will contain executables from the following examples. The \heplike package comes also with test procedures for each of the classes. To perform the tests user has to perform the command:

\begin{lstlisting}[escapeinside={(*}{*)}]
  ctest
\end{lstlisting}
or an equivalent:
\begin{lstlisting}[escapeinside={(*}{*)},]
  make test
\end{lstlisting}

\noindent If the \heplike was successfully installed the output will look as following:
\begin{lstlisting}[escapeinside={(*}{*)}]
  Test project /storage/github/HEPLike/build
  Start 1: HL_Test_YAML
  1/7 Test #1: HL_Test_YAML .....................   Passed    0.01 sec
  Start 2: HL_Limit
  2/7 Test #2: HL_Limit .........................   Passed    0.27 sec
  Start 3: HL_Br_example
  3/7 Test #3: HL_Br_example ....................   Passed    0.02 sec
  Start 4: HL_BrBifurGaussian_example
  4/7 Test #4: HL_BrBifurGaussian_example .......   Passed    0.01 sec
  Start 5: HL_Ndim_Gaussian
  5/7 Test #5: HL_Ndim_Gaussian .................   Passed    0.01 sec
  Start 6: HL_ProfLikelihood_example
  6/7 Test #6: HL_ProfLikelihood_example ........   Passed    0.25 sec
  Start 7: HL_Ndim_BifurGaussian_example
  7/7 Test #7: HL_Ndim_BifurGaussian_example ....   Passed    0.01 sec

  100% tests passed, 0 tests failed out of 7

  Total Test time (real) =   0.57 sec
\end{lstlisting}

\subsection{Available measurement}
\label{sec:datasets}

The \yaml files that contain the stored measurements are located in a second independent repository. The reason for this separation is that the \yaml files are expected to be updated more frequently then the code itself. It is expected that users and experiments will contribute to this repository. By implementing such model it is ensured that the repository will contain the most up to date measurements.

The repository can be found at: \url{https://github.com/mchrzasz/HEPLikeData}. The repository should be downloaded or cloned:
\begin{lstlisting}[escapeinside={(*}{*)}, numbers=right]
cd <some new dir>
git clone  https://github.com/mchrzasz/HEPLikeData.git       
\end{lstlisting} 

Since the repository contains only \yaml files there is no need for any compilation. The repository contains a directory \texttt{data}, where all the \yaml files are kept. It should be linked by a symbolic link to the \heplike package. Inside the \texttt{data} the measurements are grouped by experiments (ex. \lhcb, \atlas, \cms, etc.). Inside the experiment directory the measurements are grouped according to type of measurement in the collaborations, for example: \texttt{RD}, \texttt{Semileptonic}, \texttt{Charmless}, \texttt{Exotica}, etc. The names of the \yaml files should be named accordingly to publication report number. For example: \texttt{CERN-EP-2018-331.yaml}. If a single publication produced more independent measurements, user might code them in the independent files and give further information at the end of the file, for example:\texttt{CERN-PH-EP-2015-314\_q2\_01\_0.98.yaml}. 

Currently we are publishing the measurements that have been used by us in other projects~\cite{Workgroup:2017myk,Athron:2017qdc,Athron:2017yua}. The list of \yaml files with the context is presented in~\tabref{tab:yaml}.

\begin{longtable}{|L{7cm}|L{6cm}|}
\caption{Functions available in the \heplike software. \label{tab:yaml}}\\
\hline
\textbf{File} & \textbf{Description} \\
\hline
\endfirsthead
\multicolumn{2}{c}%
{\tablename\ \thetable\ -- \textit{Continued from previous page}} \\
\hline
\textbf{File} & \textbf{Description} \\
\hline
\endhead
\endfoot
\hline
  \texttt{CERN-EP-2017-100.yaml} & \yaml file encoding the measurement of branching fraction of the $\PBd \to \mu \mu$ and $\PBs \to \mu \mu$ decays~\cite{Aaij:2017vad}. \\  \hline
   \texttt{PH-EP-2015-314\_q2\_0.1\_0.98.yaml}
\texttt{PH-EP-2015-314\_q2\_11.0\_12.5.yaml}
\texttt{PH-EP-2015-314\_q2\_1.1\_2.5.yaml}
\texttt{PH-EP-2015-314\_q2\_15.0\_19.yaml}
\texttt{PH-EP-2015-314\_q2\_2.5\_4.0.yaml}
\texttt{PH-EP-2015-314\_q2\_4.0\_6.0.yaml}
\texttt{PH-EP-2015-314\_q2\_6.0\_8.0.yaml}
& \yaml files encoding the measurements of the angular coefficients of $\PBd \to \PKstar \mu \mu$ decay in different $q^2$ regions~\cite{Aaij:2015oid}. \\  \hline
   \texttt{CERN-EP-2016-141\_q2\_0.1\_0.98.yaml}
\texttt{CERN-EP-2016-141\_q2\_11.0\_12.5.yaml}
\texttt{CERN-EP-2016-141\_q2\_1.1\_2.5.yaml}
\texttt{CERN-EP-2016-141\_q2\_15.0\_19.yaml}
\texttt{CERN-EP-2016-141\_q2\_2.5\_4.0.yaml}
\texttt{CERN-EP-2016-141\_q2\_4.0\_6.0.yaml}
\texttt{CERN-EP-2016-141\_q2\_6.0\_8.0.yaml}
& \yaml files encoding the measurements of the branching fraction of the $\PBd \to \PKstar \mu \mu$ decay in different $q^2$ regions~\cite{Aaij:2016flj}. \\  \hline
   \texttt{CERN-EP-2016-215\_q2\_0.1\_0.98.yaml}
\texttt{CERN-EP-2016-215\_q2\_1.1\_2.5.yaml}
\texttt{CERN-EP-2016-215\_q2\_2.5\_4.yaml}
\texttt{CERN-EP-2016-215\_q2\_4\_6.yaml}
\texttt{CERN-EP-2016-215\_q2\_6\_8.yaml}
& \yaml files encoding the measurements of the branching fraction of the $\PBd \to \PK \Ppi \mu \mu$ decay in different $q^2$ regions~\cite{Aaij:2016kqt}. \\  \hline
\texttt{CERN-PH-EP-2015-145\_0.1\_2.yaml}
\texttt{CERN-PH-EP-2015-145\_11\_12.5.yaml}
\texttt{CERN-PH-EP-2015-145\_15\_19.yaml}
\texttt{CERN-PH-EP-2015-145\_1\_6.yaml}
\texttt{CERN-PH-EP-2015-145\_2\_5.yaml}
\texttt{CERN-PH-EP-2015-145\_5\_8.yaml}
& \yaml files encoding the measurements of the branching fraction of the $\PBs \to \Pphi \mu \mu$ decay in different $q^2$ regions~\cite{Aaij:2016kqt}. \\  \hline
\texttt{CERN-EP-2019-043.yaml} & \yaml file encoding the measurement of the $R_K$~\cite{Aaij:2019wad}. \\  \hline
\texttt{CERN-EP-2017-100\_q2\_0.045\_1.1.yaml} \texttt{CERN-EP-2017-100\_q2\_1.1\_6.yaml} & \yaml file encoding the measurement of the $R_{\PKstar}$~\cite{Aaij:2017vbb}. \\  \hline
\texttt{b2sgamma.yaml} & \yaml file encoding the HFLAV average of the $\Pbeauty \to \Pstrange \Pphoton$~\cite{Amhis:2016xyh}. \\  \hline
\texttt{RD\_RDstar.yaml} & \yaml file encoding the HFLAV average of the $R(\PD)$ and $R(\PDstar)$~\cite{Amhis:2016xyh}. \\  \hline
\texttt{HFLAV\_2016\_157.yaml}
\texttt{HFLAV\_2016\_160.yaml}
\texttt{HFLAV\_2016\_161.yaml}
\texttt{HFLAV\_2016\_162.yaml}
\texttt{HFLAV\_2016\_164.yaml}
\texttt{HFLAV\_2016\_165.yaml}
\texttt{HFLAV\_2016\_166.yaml}
\texttt{HFLAV\_2016\_167.yaml}
\texttt{HFLAV\_2016\_168.yaml}
\texttt{HFLAV\_2016\_169.yaml}
\texttt{HFLAV\_2016\_170.yaml}
\texttt{HFLAV\_2016\_171.yaml}
\texttt{HFLAV\_2016\_176.yaml}
\texttt{HFLAV\_2016\_177.yaml}
\texttt{HFLAV\_2016\_178.yaml}
\texttt{HFLAV\_2016\_179.yaml}
\texttt{HFLAV\_2016\_180.yaml}
\texttt{HFLAV\_2016\_181.yaml}
\texttt{HFLAV\_2016\_182.yaml}
\texttt{HFLAV\_2016\_183.yaml}
\texttt{HFLAV\_2016\_211.yaml}
\texttt{HFLAV\_2016\_212.yaml}
& \yaml files encoding the upper limits of $\tau$ Lepton Flavour Violation decays~\cite{Aaij:2016kqt}. \\  \hline
\end{longtable}

As already mentioned the measurements are constantly growing and there is expected that the community will contribute to develop this repository. When a new \yaml file is wrote before merging it with the repository it should be checked if it contains all the necessary information. It can be checked with the \texttt{Test\_YAML.cc} program. It can be used in the following way:

\begin{lstlisting}[escapeinside={(*}{*)}, numbers=right]
cd HEPLike
./build/Test_YAML <PATH_TO_YAML>
\end{lstlisting} 

If an entry is missing the user will be notified by a printout. The HEPLikeData repository contains also a template \yaml (data/template.yaml) file which can be used to create new measurements \yaml files.

As already mentioned we provide useful utilities for the encoded measurements. The first is the ability to create \bib file for the measurements that have been used. The user should store the \bib items or \yaml file names:

\begin{lstlisting}[escapeinside={(*}{*)}, numbers=right]
Aaij:2017vbb 
b2mumu.yaml  
\end{lstlisting} 

To prepare the \bib file user should run the \texttt{make\_citations.py} script located in the \texttt{utils} directory:
\begin{lstlisting}[escapeinside={(*}{*)}, numbers=right]
cd utils
python make_citations.py list.txt
\end{lstlisting} 
after this command a new file \texttt{references.bib}, will be created, which will contain the full \bib entries. This can be directly used in preparing the publication.

Another useful feature of \heplike is the ability to search the measurement database for relevant measurements. The script allowing for that utility is also located in the \texttt{utils}. Currently the database can be searched for using the year of publication, Arxiv number, author of the \yaml file or the unique name of the measurements. The syntax for running a search is the following:
\begin{lstlisting}[escapeinside={(*}{*)}, numbers=right]
python lookup.py --Arxiv 1705.05802
Found files:
../data/examples/RKstar_lowq2.yaml
\end{lstlisting} 

To see all available search options in the following script user can run it with help option: \texttt{python lookup.py -h}.

\section{Summary}

We have presented a computer program \heplike that enables to construct and evaluate experimental likelihoods. The software is designed to handle the interpretation of wide range of published results. It also allows to perform direct fits to data once it is provided by the experimental collaborations.

The program can be easily interfaced with other computer programs and is aimed to help users, who perform fits to experimental results in their scientific work. It is especially useful for large fitting collaborations, which till now had to implement the experimental measurements on their own. The measurement themselves are stored in \yaml files in separate repository. This allows for easy extensions of the database without the need of compilation. Furthermore, users and experimental collaborations can share their encoded measurements with the community.

\section*{Acknowledgments}

This work is partly supported by the CERN FCC Design Study Program. The research of M.~Chrzaszcz is funded by the Polish National Agency for Academic Exchange under the Bekker program. M.~Chrzaszcz is also grateful to Foundation for Polish Science (FNP) for its support.

We would like thank Mike Williams, Patrick Koppenburg, Pat Scott, Danny van Dyk and Maria Moreno Llacer for invaluable comments about our manuscript.





\bibliographystyle{elsarticle-num}
\bibliography{references}







\end{document}